\documentclass{pasj00}

\SetRunningHead{Matsushita et al.}
	{Dense and Warm Molecular Gas and Warm Dust in Nearby Galaxies}

\Received{}
\Accepted{}
\Published{}

\begin{document}

\title{Dense and Warm Molecular Gas and Warm Dust in Nearby Galaxies}

\author{Satoki \textsc{Matsushita}}
\affil{Institute of Astronomy and Astrophysics,
	Academia Sinica, P.O.\ Box 23-141, Taipei 10617, Taiwan, R.O.C.}
\email{satoki@asiaa.sinica.edu.tw}
\author{Ryohei \textsc{Kawabe}}
\affil{Nobeyama Radio Observatory, Minamimaki, Minamisaku, Nagano,
	384-1305}
\author{Kotaro \textsc{Kohno}}
\affil{Institute of Astronomy, University of Tokyo,
	2-21-1 Osawa, Mitaka, Tokyo 181-0015}
\author{Tomoka \textsc{Tosaki}}
\affil{Department of Geoscience, Joetsu University of Education,
	1 Yamayashiki, Joetsu, Niigata, 043-8512}
\and
\author{Baltasar \textsc{Vila-Vilar\'o}}
\affil{ALMA Santiago / ESO,
	Av Apoquindo 3846, Piso 19, Las Condes, Santiago, Chile}


\maketitle

\begin{abstract}
We performed $^{12}$CO(1 -- 0), $^{13}$CO(1 -- 0), and HCN(1 -- 0)
single-dish observations (beam size $\sim14\arcsec-18\arcsec$) toward
nearby starburst and non-starburst galaxies using the Nobeyama 45~m
telescope.
The $^{13}$CO(1 -- 0) and HCN(1 -- 0) emissions were detected from
all the seven starburst galaxies, with the intensities of both lines
being similar (i.e., the ratios are around unity).
On the other hand, for case of the non-starburst galaxies, the
$^{13}$CO(1 -- 0) emission was detected from all three galaxies,
while the HCN(1 -- 0) emission was weakly or not detected in past
observations.
This result indicates that the HCN/$^{13}$CO intensity ratios are
significantly larger ($\sim1.15\pm0.32$) in the starburst galaxy
samples than the non-starburst galaxy samples ($<0.31\pm0.14$).
The large-velocity-gradient model suggests that the molecular gas in
the starburst galaxies have warmer and denser conditions than that in
the non-starburst galaxies, and the photon-dominated-region model
suggests that the denser molecular gas is irradiated by stronger
interstellar radiation field in the starburst galaxies than that in
the non-starburst galaxies.
In addition, HCN/$^{13}$CO in our sample galaxies exhibit strong
correlations with the IRAS $25~\micron$ flux ratios.
It is a well established fact that there exists a strong correlation
between dense molecular gas and star formation activities, but our
results suggest that molecular gas temperature is also an important
parameter.
\end{abstract}

\section{Introduction}
\label{sect-intro}

Galaxies consist of vast numbers of stars formed from molecular
clouds.
It has therefore been suggested that the process of star formation
in galaxies depends largely on the distribution, kinematics, and
physical conditions of the molecular gas.
One of the primary molecular gas properties critical to star
formation is the number density.
Observational studies of molecular clouds in our Galaxy suggest that
stars are formed from the dense locale of molecular clouds, rather
than the diffuse regions (e.g., \cite{lad92}).
Extragalactic single-dish molecular gas observations indicate that
the overall amount of dense molecular gas, which can be traced by the
HCN(1 -- 0) luminosity, is tightly correlated with the global star
formation rate, which can in turn be traced by the far-infrared (FIR)
luminosity observed with the Infrared Astronomical Satellite (IRAS)
\citep{sol92,gao04a,gao04b}.
This correlation is much tighter than that between the diffuse
molecular gas, or in other words, the total amount of molecular gas
that can be traced by $^{12}$CO(1 -- 0), and the global star
formation rate (e.g., \cite{san91,you91,you96,gao04b}).
In addition, \citet{sol92} and \citet{gao04b} also raise the
possibility of the fact that the HCN/$^{12}$CO luminosity ratio
correlates well with the FIR/$^{12}$CO luminosity ratio, namely the
fraction of dense molecular gas correlates well with the star
formation efficiency, and this suggests that HCN/$^{12}$CO is a good
indicator for star formation.

High spatial resolution interferometric observations toward the
center of a galaxy with a high infrared luminosity indicate that
extragalactic star forming regions coincide well with the
distributions of dense molecular gas (HCN) \citep{koh99}.
The distributions of diffuse molecular gas ($^{12}$CO), on the other
hand, only loosely match those of star forming regions.
As a matter of fact, the HCN/$^{12}$CO integrated intensity ratios
are high ($\sim0.1-0.2$) in active star forming regions in the
centers of galaxies with high infrared luminosity
($>10^{10}$ L$_{\odot}$), but low ($<0.1$) in the centers of galaxies
with low star forming activities or post-starburst activities
\citep{koh99,rey99,koh02}.
Star forming regions in the outer parts of galaxies, where normally
have low infrared luminosities or low star formation rate,
have low HCN/$^{12}$CO at giant molecular cloud scale
\citep{hel97,bro05}.
These results strongly imply that the density of molecular gas play a
critical role in the active star formation phenomena.

The relation between star formation and temperature, another
important molecular gas property, has so far not been well
established.
There are several surveys observing $^{12}$CO(2 -- 1) and
$^{12}$CO(1 -- 0) lines toward galaxies with various star formation
activities (such as starburst, interacting, and quiescent galaxies),
and most of the surveys yield an average $^{12}$CO(2 -- 1)/(1 -- 0)
of around unity \citep{bra93,aal95,haf03}.
Surveys focusing on the $^{12}$CO(3 -- 2) and $^{12}$CO(1 -- 0) lines
indicate that variations in $^{12}$CO(3 -- 2)/(1 -- 0) tend to have
somewhat larger dependence on star formation activities as compared
to $^{12}$CO(2 -- 1)/(1 -- 0); starburst galaxies tend to have higher
$^{12}$CO(3 -- 2)/(1 -- 0) of around unity, but normal galaxies tend
to have lower ratio of around 0.5 or lower \citep{haf03,dum01,dev94}.
These multi-transition $^{12}$CO observations suggest that the
$^{12}$CO lines in nearby normal and starburst galaxies are optically
thick, with a tendency to be saturated to unity, namely thermalized,
in the vicinity of certain star formation activities.
It is therefore difficult to deduce correlations between star
formation activities and multi-transition $^{12}$CO ratios.

The intensities of the $^{13}$CO(1 -- 0) lines, on the other hand,
display more drastic changes than those of the $^{12}$CO lines with
respect to star formation activities.
In merging galaxies, the $^{13}$CO(1 -- 0) intensities tend to be
very low, and $^{12}$CO(1 -- 0)/$^{13}$CO(1 -- 0) often exceed 20,
in contrast to normal or starburst galaxies (not including extreme
starbursts) that have an average value of $\sim10$
\citep{aal91,cas92,aal95}.
However, this clear contrast in $^{12}$CO(1 -- 0)/$^{13}$CO(1 -- 0)
is only manifested between merging galaxies and normal/starburst
galaxies, and there exists only a weak trend between normal and
starburst galaxies as a function of dust temperature or the 60~$\mu$m
and 100~$\mu$m FIR flux ratio \citep{sag91,aal91,aal95}.

In the study of M51 multi-line observations, we discovered an
excellent density and temperature probe, the
HCN(1 -- 0)/$^{13}$CO(1 -- 0) ratio
(\cite{mat98}, hereafter Paper I; \cite{mat99}).
The utility of this ratio has been confirmed by higher-J
interferometric $^{12}$CO observations \citep{mat04}.

In this paper, we present HCN(1 -- 0), $^{13}$CO(1 -- 0), and
$^{12}$CO(1 -- 0) observations toward galaxies with various levels of
star formation activities, and discuss the relation between the
physical conditions of molecular gas and the levels of star formation
activities.
In addition, we compare the IR properties of these galaxies, and show
that the molecular gas and dust are localized in similar regions and
strongly irradiated by the radiation from starburst regions.

\begin{figure}
\begin{center}
\FigureFile(80mm,80mm){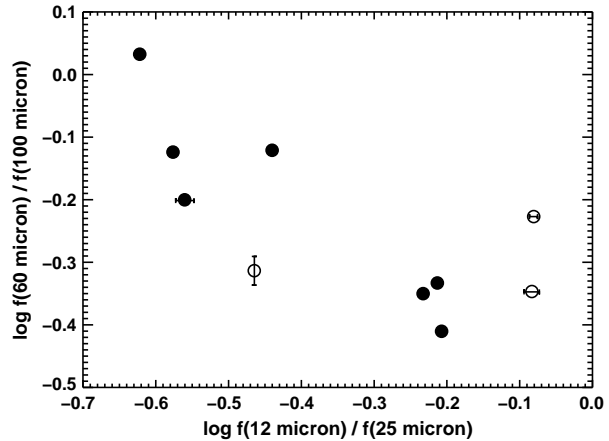}
\end{center}
\caption{Correlation diagrams between the
	$f(12~\micron)/f(25~\micron)$ and the
	$f(60~\micron)/f(100~\micron)$ IR flux ratios.
	Filled and open circles denote the starburst and the
	non-starburst galaxy samples, respectively.
	Our sample galaxies extend over a wide range of IR flux ratios
	reported in \citet{hel86} and \citet{dal01}, covering a wide
	spectrum of star formation activities.
\label{fig-sample-ir}}
\end{figure}

\begin{table*}
\begin{center}
\caption{Basic parameters of the starburst galaxy samples.}
\label{table-sample-sb}
\begin{tabular}{llllcrcc}
\hline
Galaxy & Type\footnotemark[$*$] & $\alpha$(B1950) & $\delta$(B1950)
	& Ref.\footnotemark[$\dagger$] & Dist.\footnotemark[$\ddagger$]
	& Ref.\footnotemark[$\S$] & $L_{\rm FIR}$\footnotemark[$\|$] \\
\hline
NGC~253  & SAB(s)c     & \timeform{00h45m05s.63}
	& \timeform{-25D33'40".5}  & 1 &  3.5 &  8 & 10.29 \\
NGC~2146 & SB(s)ab pec & \timeform{06h10m41s.10}
	& \timeform{+78D22'28".1}  & 2 & 17.2 &  9 & 10.93 \\
NGC~2903 & SAB(rs)bc   & \timeform{09h29m20s.26}
	& \timeform{+21D43'20".74} & 3 &  8.9 & 10 & 10.05 \\
M82      & I0 sp       & \timeform{09h51m43s.5}
	& \timeform{+69D55'00".0}  & 4 &  3.9 & 11 & 10.61 \\
NGC~3504 & (R)SAB(s)ab & \timeform{11h00m28s.53}
	& \timeform{+28D14'31".2}  & 5 & 26.5 &  9 & 10.56 \\
NGC~6946 & SAB(rs)cd   & \timeform{20h33m49s.2}
	& \timeform{+59D58'49".5}  & 6 &  5.9 & 12 & 10.01 \\
NGC~6951 & SAB(rs)bc   & \timeform{20h36m36s.59}
	& \timeform{+65D55'46".0}  & 7 & 26.8 & 13 & 10.46 \\
\hline
\multicolumn{7}{@{}l@{}}{\hbox to 0pt{\parbox{140mm}{\footnotesize
	Notes.
	\par\noindent
	\footnotemark[$*$] Morphological type taken from RC3.
	\par\noindent
	\footnotemark[$\dagger$] References for the positions.
	\par\noindent
	\footnotemark[$\ddagger$] Distance in Mpc.
	\par\noindent
	\footnotemark[$\S$] References for distance.
	\par\noindent
	\footnotemark[$\|$] Far infrared luminosity taken from
		\citet{san03}.
	\par\noindent
	References --- (1) \citet{ket93}. (2) \citet{con82}.
		 (3) 2.2~$\mu$m peak of \citet{wyn85}.   (4) \citet{she95}.
		 (5) \citet{con90}.  (6) \citet{tur83}.  (7) \citet{sai94}.
		 (8) \citet{rek05}.  (9) \citet{tul88}. (10) \citet{dro00}.
		(11) \citet{sak99}. (12) \citet{kar00}. (13) \citet{val03}.
	}\hss}}
\end{tabular}
\end{center}
\end{table*}

\section{Sample Galaxies}
\label{sect-sample}

For this survey, we picked 7 starburst galaxies and 3 non-starburst
galaxies so as to cover a wide range of star formation activities.
We summarized the basic parameters of the starburst and non-starburst
galaxy samples in Table~\ref{table-sample-sb} and
Table~\ref{table-sample-npb}, respectively.
To ensure the wide spectrum of star formation activities of our
samples, we plot the IR flux ratios between $12~\mu$m and $25~\mu$m,
$f(12~\micron)/f(25~\micron)$, and that between $60~\mu$m and
$100~\mu$m, $f(60~\micron)/f(100~\micron)$ in Fig.~\ref{fig-sample-ir}.
It has well established that this correlation traces star formation
activities \citep{hel86,dal01}.
Our sample galaxies extend over a wide range of the IR flux ratios,
covering most of the range of the diagrams shown in \citet{hel86} and
\citet{dal01}, thus ensuring the wide coverage of star formation
activities.

\subsection{Starburst Galaxy Sample}
\label{sect-sb}

Among the starburst galaxy samples are M82, NGC~253, NGC~2146,
NGC~3504, NGC~2903, NGC~6946, and NGC~6951, whose FIR luminosities
are on the order of $10^{10}$~L$_{\odot}$ \citep{san03}.

M82 and NGC~253 exhibit energetic phenomena at various wavelengths,
such as H$\alpha$, X-ray, and molecular gas outflows (e.g.,
\cite{str04,nak87}) and molecular superbubbles (e.g.,
\cite{mat05,sak06}).
These galaxies are therefore known as prototypical starburst galaxies
(e.g., \cite{rie80,rie88a,dev89}).

NGC~2146 and NGC~3504 display star formation activities comparable to
M82 and NGC~253 on the basis of 1.65, 2.2, and 10 $\mu$m photometric
observations \citep{dev89,dev94}.
In particular, NGC~2146 hosts a kpc-scale starburst-driven X-ray
outflow \citep{arm95,del99,inu05} and molecular outflow and
superbubbles \citep{tsa09} from the central region, which are similar
to those in M82 or NGC 253, and therefore strongly support the
existence of ongoing starburst.

NGC~2903 is one of the S\'ersic-Pastoriza galaxies (a catalogue of
galaxies with peculiar or complex nuclei such as ``hot spots'' or
``amorphous nucleus''; \cite{ser65,ser67,ser73}) and its nuclear
appearance is classified as ``hot spots''.
This nuclear ``hot spots'' region is a powerful IR source, and it has
been proposed that this phenomenon could be the result of a starburst
\citep{rie80,tel80}, a conjecture that has been supported by radio
and IR observations \citep{wyn85,alo01}.

NGC~6946 is a large spiral galaxy with a small nuclear bar
\citep{elm98} and ongoing nuclear starburst \citep{tur83,eng96}.

NGC~6951 is also a S\'ersic-Pastoriza galaxy, with a circumnuclear
``hot spots'' ring \citep{bar95,elm99}.
This ring is bright in H$\alpha$ \citep{woz95,gon97} and exhibits
high star formation rate (SFR) of $\sim3-4$ M$_{\odot}$~yr$^{-1}$ as
well as high star formation efficiency (SFE), comparable to those in
the central regions of nearby starburst galaxies \citep{koh99,woz95}.

\begin{table*}
\begin{center}
\caption{Basic parameters of the non-starburst galaxy samples.}
\label{table-sample-npb}
\begin{tabular}{llllccccl}
\hline
Galaxy & Type\footnotemark[$*$] & $\alpha$(B1950) & $\delta$(B1950)
	& Ref.\footnotemark[$\dagger$] & Dist.\footnotemark[$\ddagger$]
	& Ref.\footnotemark[$\S$] & Activity$^{\|}$
	& $L_{\rm FIR}$\footnotemark[$\#$] \\
\hline
NGC~4736 & (R)SA(r)ab  & \timeform{12h48m31s.910}
	& \timeform{+41D23'31".78} & 1 &  4.3 & 3 & Post-SB & 9.59  \\
NGC~4826 & (R)SA(rs)ab & \timeform{12h35m16s.07}
	& \timeform{+21D57'13".5}  & 2 &  4.1 & 3 &    T2   & 8.98  \\
NGC~5195 & I0 pec      & \timeform{13h27m53s.27}
	& \timeform{+47D31'25".5}  & 2 &  8.4 & 4 & Post-SB & 9.44: \\
\hline
\multicolumn{8}{@{}l@{}}{\hbox to 0pt{\parbox{160mm}{\footnotesize
	Notes.
	\par\noindent
	\footnotemark[$*$] Morphological type taken from RC3.
	\par\noindent
	\footnotemark[$\dagger$] References for the positions.
	\par\noindent
	\footnotemark[$\ddagger$] Distance in Mpc.
	\par\noindent
	\footnotemark[$\S$] References for distance.
	\par\noindent
	\footnotemark[$\|$] Star formation activity. ``SB'' and ``T2''
		denote starburst and type 2 transition object,
		respectively.
		See text for details.
	\par\noindent
	\footnotemark[$\#$] Far infrared luminosity taken from
		\citet{san03}.
		A value with a colon (``:'') indicates that the value has a
		large uncertainty.
	\par\noindent
	References --- (1) \citet{tur94}. (2) \citet{hum87}.
		(3) \citet{tul88}. (4) \citet{fel97}.
	}\hss}}
\end{tabular}
\end{center}
\end{table*}

\subsection{Non-Starburst Galaxy Sample}
\label{sect-nsb}

The non-starburst galaxy sample consists of NGC 4736, NGC 4826, and
NGC 5195, whose FIR luminosities are around $10^{9}$~L$_{\odot}$
\citep{san03}.

The optical spectra of the nuclear regions of both NGC~4736 and
NGC~5195 exhibit strong Balmer absorption lines
\citep{hof95,tan96,sau96,gre98}.
Since these strong Balmer absorption lines are typically found in the
spectra of A-type stars, stellar populations of galaxies with such
absorption lines are believed to be dominated by A-type stars.
Indeed, the stellar population synthesis analysis implies that the
dominant stellar populations of NGC~4736 and NGC~5195 extend up to at
most A4 to A7, and that the optical light is dominated by these
stellar populations \citep{pri77,war74}.
The likely mechanism for the creation of ``A-type star dominated''
galaxies is as follows:
Starburst phenomena produce enormous numbers of early to late-type
stars, assuming a general initial mass function.
Early-type stars (OB stars) live at most $10^{8}$ years, so that at
the time $10^{9}$ years after the beginning of the starburst, the
early-type stars are burned out, with vast numbers of A-type stars
remaining.
Therefore, galaxies with numerous A-type stars can be considered
post-starburst galaxies \citep{rie88b,rie88a,wal88}.
Molecular gas observations support the hypothesis that, in the Balmer
absorption dominated regions, there is little molecular gas
\citep{tos97}.
In some cases, molecular gas exists in the ``A-type star dominated''
galaxies, but it is stable against gravitational instability
\citep{shi98,koh02} with almost no high-density gas present
\citep{koh98,koh02}.
Therefore, we classified NGC~4736 and NGC~5195 as ``post-starburst''
as in other articles.

NGC~4826 is classified as a type 2 transition object (the
[O\emissiontype{I}] line from the nucleus has a strength intermediate
between those of H\emissiontype{II} nuclei and LINERs; \cite{hof97}).
Even if all of the H$\alpha$ emission of this galaxy originates in
the star forming regions, the extinction corrected total H$\alpha$
luminosity, $L({\rm H}\alpha)$, of $5.7\times10^{40}$~erg~s$^{-1}$
\citep{you96} indicates a SFR of only
$\sim0.45$~M$_{\odot}$~yr$^{-1}$, using the relationship between the
SFR and the $L({\rm H}\alpha)$ \citep{ken98}.
Hence this galaxy can be safely regarded as a non-starburst galaxy.

\section{Observations}
\label{sect-obs}

Observations of the central regions of the sample galaxies using the
Nobeyama 45~m telescope were carried out in the HCN(1 -- 0),
$^{13}$CO(1 -- 0), and $^{12}$CO(1 -- 0) lines (rest frequency =
86.632~GHz, 110.201~GHz, and 115.271~GHz, respectively).
We observed $^{13}$CO(1 -- 0) and HCN(1 -- 0) simultaneously for
the starburst samples during 1997 February -- April, 1997 December,
and 1999 March -- April (the 1999 observation run was only for
NGC~6951).
The $^{13}$CO(1 -- 0) and $^{12}$CO(1 -- 0) simultaneous observations
for the non-starburst samples and M82 were performed during 1998
March -- April.

Two SIS receivers, S100 and S80, were used, and these were tuned to
the $^{13}$CO(1 -- 0) and HCN(1 -- 0) lines, respectively, for the
$^{13}$CO and HCN simultaneous observations.
The half power beam widths (HPBW) for the $^{13}$CO and HCN
simultaneous observations (except for NGC~6951) were $15\arcsec$ and
$17\arcsec$, and the main-beam efficiencies, $\eta_{\rm mb}$, were
$0.51\pm0.03$ and $0.47\pm0.02$, respectively.
In case of NGC 253, we applied $\eta_{\rm mb}$ of 0.48 and 0.35,
respectively, due to the low declination (and therefore the low
elevation at the telescope site) of this source.
The HPBW for the NGC~6951 observations were $15\arcsec$ and
$18\arcsec$, and $\eta_{\rm mb}$ were $0.48\pm0.02$ and
$0.49\pm0.02$, respectively.
The differences in the HPBW and $\eta_{\rm mb}$ from other
observations are due to the different observation year for this
galaxy.

For the $^{12}$CO(1 -- 0) and $^{13}$CO(1 -- 0) simultaneous
observations, S100 and S80 were tuned to $^{12}$CO and $^{13}$CO,
respectively.
Since the different (S80 instead of S100) receiver was used for the
$^{13}$CO line, the HPBW and $\eta_{\rm mb}$ were different from the
$^{13}$CO and HCN simultaneous observations;
the HPBW for the $^{13}$CO and $^{12}$CO simultaneous observations
were both 14\arcsec, and the $\eta_{\rm mb}$ were $0.40\pm0.03$ and
$0.51\pm0.03$, respectively.

The telescope pointing was checked and corrected every hour, and the
absolute pointing errors were less than 4\arcsec\ during the
observations.
Due to the use of two SIS receivers, there were no relative pointing
offsets between two simultaneously observed lines.
The single side-band system noise temperatures at 87~GHz, 110~GHz,
and 115~GHz throughout the observations were 200 -- 400~K,
300 -- 1400~K, and 600 -- 1800~K, respectively.
As back-ends, we used 2048-channel wide band acousto-optical
spectrometers (AOS), with a total bandwidth of 250~MHz, corresponds
to 650~km~s$^{-1}$ at the $^{12}$CO(1 -- 0) frequency.
The line intensity calibration was accomplished by the chopper-wheel
method \citep{uli76}, and the relative intensity errors were
calibrated at every observation with the intensity of W51 main for
NGC~6951 and IRC$+$10216 for the rest of the observations.

\begin{figure*}
\begin{center}
\FigureFile(170mm,170mm){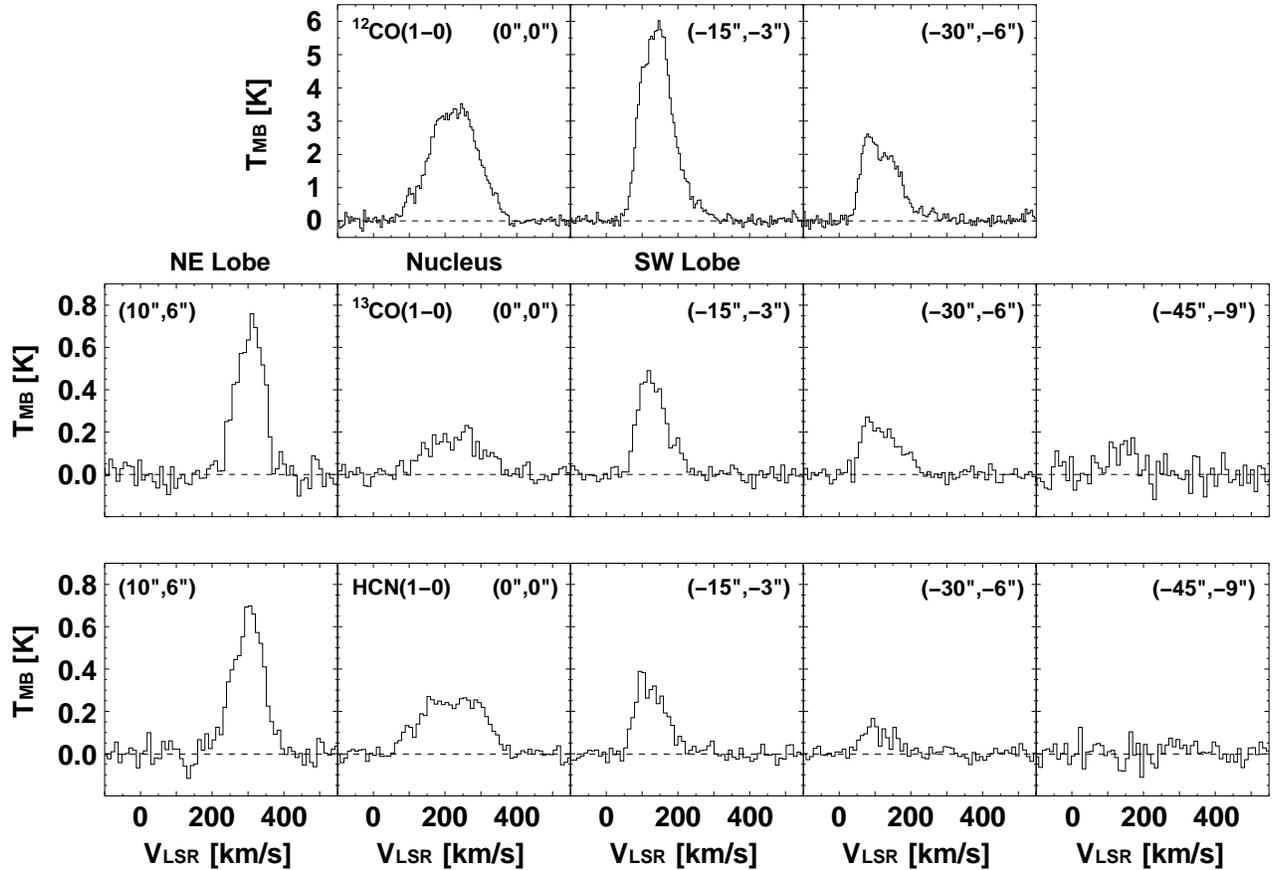}
\end{center}
\caption{$^{12}$CO(1 -- 0) (top row), $^{13}$CO(1 -- 0) (middle row),
	and HCN(1 -- 0) (bottom row) spectra of the central region of
	the prototypical starburst galaxy M82.
	The horizontal axis is the LSR velocity, $V_{\rm LSR}$, and the
	vertical axis is the main beam temperature, $T_{\rm MB}$.
	The reference point (\timeform{0"}, \timeform{0"}) is the
	position of the galactic nucleus as determined from the peak of
	the 2.2~$\mu$m source (see Table~\ref{table-sample-sb}).
	The other points are in R.A.\ and Decl.\ offset from the
	reference point.
\label{fig-res-m82}}
\end{figure*}

We observed the central regions of sample galaxies pointed toward the
nuclei except for M82.
M82 was observed toward five points for the $^{13}$CO(1 -- 0) and
HCN(1 -- 0) lines, and three points for the $^{12}$CO(1 -- 0) line.
The observed positions are shown in Fig.~\ref{fig-res-m82} and
Table~\ref{table-res-prop}.
The positions (\timeform{0"}, \timeform{0"}), (\timeform{-15"},
\timeform{-3"}), and (\timeform{10"}, \timeform{6"}) are the
so-called the nucleus, the South-West (SW) lobe, and the North-East
(NE) lobe, respectively (e.g., \cite{she95}).

The data were reduced using the reduction software package NEWSTAR of
the Nobeyama Radio Observatory.
The baselines of raw spectra were subtracted by fitting linear lines,
or if necessary, second order polynomials.
The line-free velocity range for each galaxy was defined with
avoiding the line emission velocity range based on the previously
published CO observations.
The raw spectra with higher order baseline fluctuation were not used
in our final spectra.

\section{Results}
\label{sect-res}

The parameters of the observed lines from all of the sample galaxies
(peak main beam brightness temperature, $T_{\rm MB}$, and integrated
intensity, $I$) are summarized in Table~\ref{table-res-prop}, and the
calculated integrated intensity ratios are summarized in
Table~\ref{table-res-ratio}.
Since we did not observe the $^{12}$CO(1 -- 0) line for the starburst
galaxy samples other than M82 and the HCN(1 -- 0) line for the
non-starburst galaxy samples, we made use instead of the data taken
with the Nobeyama 45~m telescope from various published papers
(see notes in Table~\ref{table-res-ratio}).

Fig.~\ref{fig-res-m82} shows the $^{12}$CO, $^{13}$CO, and HCN
line spectra toward the observed positions of the central region of
M82, one of the starburst galaxy samples.
The overall line shapes of each positions are similar across these
three line spectra.
The line shapes of all three spectra at the nucleus indicate constant
brightness temperatures around the line center,
$V_{\rm LSR}\sim150-300$~km~s$^{-1}$.
It is also obvious that the $^{13}$CO line is weaker than other two
lines.
On the other hand, the line shapes at the off-center indicate peak
structures, and the peak brightness temperatures of the NE and SW
lobes exceed those of the nucleus in all the observed lines.
We did not detect the HCN line in the farthest observed region,
(\timeform{-45"}, \timeform{-9"}).
The line intensity ratios are, on the other hand, different from the
brightness temperature distributions.
At the center of M82, the intensity ratios (HCN/$^{13}$CO,
$^{12}$CO/$^{13}$CO, and HCN/$^{12}$CO) are all high, but these
ratios gradually decrease with increasing offsets from the center
(Table~\ref{table-res-ratio}).

\begin{figure*}
\begin{center}
\FigureFile(170mm,170mm){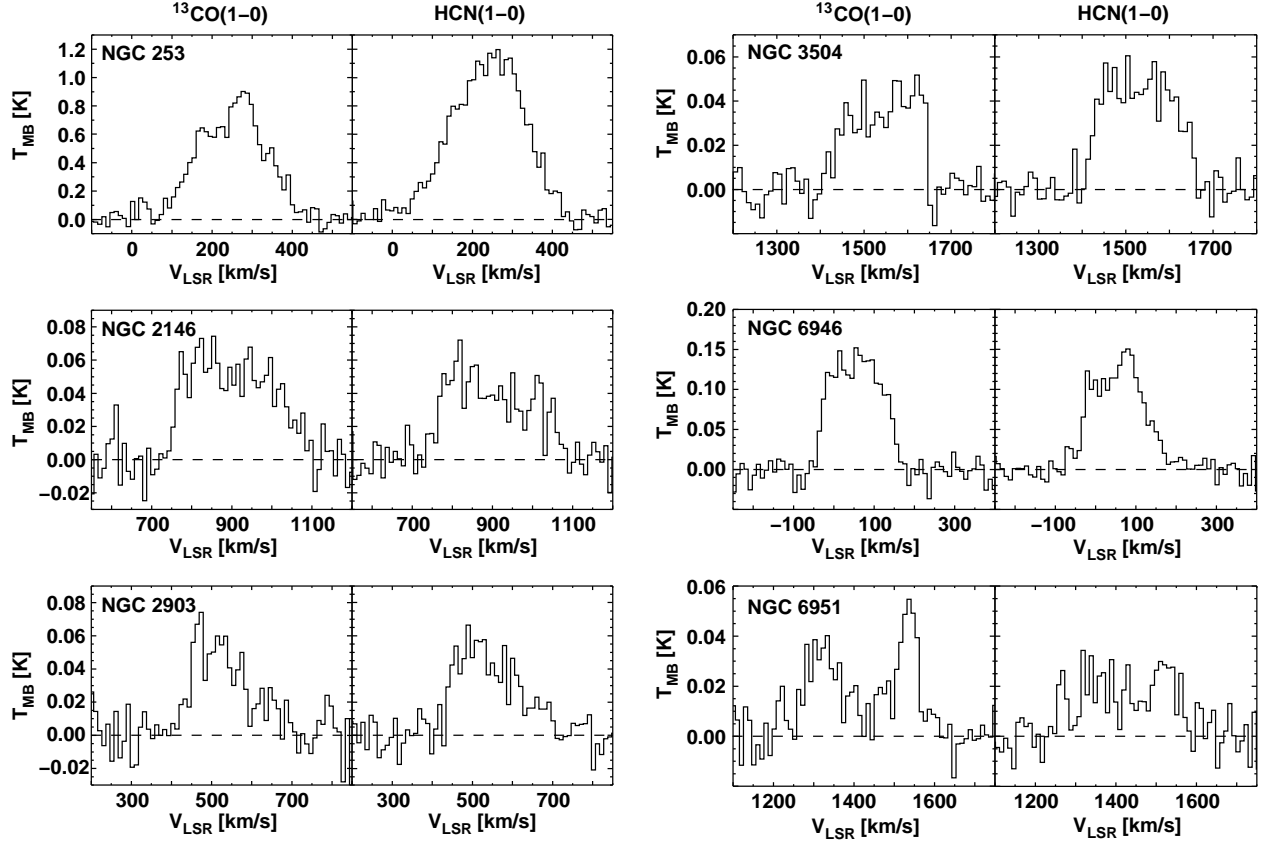}
\end{center}
\caption{$^{13}$CO(1 -- 0) and HCN(1 -- 0) spectra toward the center
	of the starburst galaxy samples.
	Two spectra are shown for each galaxy; the spectrum on the
	left-hand side denotes the $^{13}$CO(1 -- 0) line, and that on
	the right-hand side denotes the HCN(1 -- 0) line.
	The horizontal and vertical axes are $V_{\rm LSR}$ and
	$T_{\rm MB}$, respectively.
	Note that the scale for the vertical axis is the same for the
	$^{13}$CO and HCN spectra.
\label{fig-res-sb}}
\end{figure*}

Fig.~\ref{fig-res-sb} shows $^{13}$CO(1 -- 0) and HCN(1 -- 0) line
spectra of the other starburst galaxy samples.
Both lines have similar line shapes and intensities for all galaxies,
which is similar to those for the central $\pm\timeform{15"}$ region
of M82.
We note again that since the HCN and $^{13}$CO lines were observed
simultaneously, there is no HCN/$^{13}$CO integrated intensity ratio
errors due to the relative pointing offsets.

\begin{table*}
\begin{center}
\caption{Observed properties for sample galaxies.}
\label{table-res-prop}
\begin{tabular}{lccccccc}
\hline
Galaxy & Position\footnotemark[$*$]
	& $T_{\rm MB}$($^{12}$CO)\footnotemark[$\dagger$]
	& $T_{\rm MB}$($^{13}$CO)\footnotemark[$\dagger$]
	& $T_{\rm MB}$(HCN)\footnotemark[$\dagger$]
	& $I$($^{12}$CO)\footnotemark[$\ddagger$]
	& $I$($^{13}$CO)\footnotemark[$\ddagger$]
	& $I$(HCN)\footnotemark[$\ddagger$] \\
\hline
NGC~253   & (\timeform{0"},   \timeform{0"})  &      ---      &  $0.43\pm0.03$
	&  $0.42\pm0.02$  &      ---     &  $168\pm15$  &  $258\pm22$  \\
NGC~2146  & (\timeform{0"},   \timeform{0"})  &      ---      & $0.074\pm0.015$
	& $0.072\pm0.007$ &      ---     & $16.0\pm1.3$ & $12.3\pm0.7$ \\
NGC~2903  & (\timeform{0"},   \timeform{0"})  &      ---      & $0.074\pm0.008$
	& $0.066\pm0.009$ &      ---     &  $9.1\pm0.7$ &  $9.8\pm0.6$ \\
M82       & (\timeform{0"},   \timeform{0"})  & $3.52\pm0.08$ &  $0.23\pm0.03$
	&  $0.27\pm0.02$  &  $565\pm33$  & $34.5\pm2.6$ & $55.4\pm2.6$ \\
(NE lobe) & (\timeform{+10"}, \timeform{+6"}) &      ---      &  $0.76\pm0.04$
	&  $0.70\pm0.04$  &      ---     & $66.1\pm4.4$ & $71.2\pm3.6$ \\
(SW lobe) & (\timeform{-15"}, \timeform{-3"}) & $6.02\pm0.13$ &  $0.49\pm0.03$
	&  $0.39\pm0.03$  &  $646\pm38$  & $45.2\pm2.9$ & $36.2\pm1.9$ \\
          & (\timeform{-30"}, \timeform{-6"}) & $2.61\pm0.33$ &  $0.27\pm0.05$
	&  $0.17\pm0.02$  &  $288\pm20$  & $29.0\pm2.8$ & $12.2\pm0.9$ \\
          & (\timeform{-45"}, \timeform{-9"}) &      ---      &  $0.17\pm0.05$
	&     $<0.08$     &      ---     & $12.7\pm2.2$ &   $<3.2$     \\
NGC~3504  & (\timeform{0"},   \timeform{0"})  &      ---      & $0.052\pm0.005$
	& $0.060\pm0.009$ &      ---     &  $7.7\pm0.5$ & $10.1\pm0.6$ \\
NGC~6946  & (\timeform{0"},   \timeform{0"})  &      ---      & $0.152\pm0.012$
	& $0.151\pm0.008$ &      ---     & $21.9\pm1.4$ & $21.4\pm1.0$ \\
NGC~6951  & (\timeform{0"},   \timeform{0"})  &      ---      & $0.055\pm0.009$
	& $0.034\pm0.007$ &      ---     &  $8.1\pm0.6$ &  $6.1\pm0.5$ \\
\hline
NGC~4736  & (\timeform{0"},   \timeform{0"})  & $0.26\pm0.01$ & $0.043\pm0.012$
	&       ---       & $44.1\pm2.7$ &  $6.3\pm1.4$ &    ---       \\
NGC~4826  & (\timeform{0"},   \timeform{0"})  & $0.53\pm0.05$ &  $0.11\pm0.03$
	&       ---       & $87.0\pm5.9$ & $17.2\pm3.6$ &    ---       \\
NGC~5195  & (\timeform{0"},   \timeform{0"})  & $0.26\pm0.04$ &  $0.10\pm0.02$
	&       ---       & $37.9\pm2.8$ & $12.0\pm2.2$ &    ---       \\
\hline
\multicolumn{8}{@{}l@{}}{\hbox to 0pt{\parbox{170mm}{\footnotesize
	Notes.
	\par\noindent
	\footnotemark[$*$] Offset from the position of the nucleus indicated
		in Tables~\ref{table-sample-sb} and \ref{table-sample-npb}.
	\par\noindent
	\footnotemark[$\dagger$] Peak main-beam temperature in K.
		Quoted uncertainties are 1$\sigma$ and upper limits are 2$\sigma$.
	\par\noindent
	\footnotemark[$\ddagger$] Integrated intensity in K~km~s$^{-1}$.
		We define the integrated intensity as $I = \int T_{\rm MB} dv$.
		Quoted uncertainties are 1$\sigma$ and upper limits are 2$\sigma$.
	}\hss}}
\end{tabular}
\end{center}
\end{table*}

\begin{table*}
\begin{center}
\caption{Integrated intensity ratios for sample galaxies.}
\label{table-res-ratio}
\begin{tabular}{lcccc}
\hline
Galaxy & Position\footnotemark[$*$]
	& HCN/$^{13}$CO\footnotemark[$\dagger$]
	& $^{12}$CO/$^{13}$CO\footnotemark[$\dagger$]
	& HCN/$^{12}$CO\footnotemark[$\dagger$] \\
\hline
NGC~253   & (\timeform{0"},   \timeform{0"})  &  $1.53\pm0.19$
	&  $7.4\pm0.6$\footnotemark[$\|$]
	& $0.209\pm0.018$\footnotemark[$\|$] \\
NGC~2146  & (\timeform{0"},   \timeform{0"})  &  $0.77\pm0.08$
	& $11.8\pm1.0$\footnotemark[$\#$]
	& $0.065\pm0.004$\footnotemark[$\#$] \\
NGC~2903  & (\timeform{0"},   \timeform{0"})  &  $1.08\pm0.11$
	& $11.3\pm0.9$\footnotemark[$**$]
	& $0.095\pm0.006$\footnotemark[$**$] \\
M82       & (\timeform{0"},   \timeform{0"})  &  $1.61\pm0.14$ & $16.4\pm1.6$
	& $0.098\pm0.007$ \\
(NE lobe) & (\timeform{+10"}, \timeform{+6"}) &  $1.08\pm0.09$ &      ---
    &       ---       \\
(SW lobe) & (\timeform{-15"}, \timeform{-3"}) &  $0.80\pm0.07$ & $14.3\pm1.2$
    & $0.056\pm0.004$ \\
          & (\timeform{-30"}, \timeform{-6"}) &  $0.42\pm0.05$ &  $9.9\pm1.2$
    & $0.042\pm0.004$ \\
          & (\timeform{-45"}, \timeform{-9"}) &  $<0.25$       &      ---
    &       ---       \\
NGC~3504  & (\timeform{0"},   \timeform{0"})  &  $1.32\pm0.12$
	&  $8.8\pm0.6$\footnotemark[$\dagger\dagger$]
	& $0.094\pm0.008$\footnotemark[$\dagger\dagger$] \\
NGC~6946  & (\timeform{0"},   \timeform{0"})  &  $0.98\pm0.08$
	&  $9.7\pm0.6$\footnotemark[$\#$]
	& $0.101\pm0.005$\footnotemark[$\#$] \\
NGC~6951  & (\timeform{0"},   \timeform{0"})  &  $0.75\pm0.08$
	&  $6.0\pm0.5$\footnotemark[$**$]
	& $0.125\pm0.010$\footnotemark[$**$] \\
\hline
NGC~4736  & (\timeform{0"},   \timeform{0"})
	&    $<0.32$\footnotemark[$\ddagger$]    &  $7.0\pm1.6$
	&  $<0.042$\footnotemark[$\ddagger$] \\
NGC~4826  & (\timeform{0"},   \timeform{0"})
	& $0.48\pm0.10$\footnotemark[$\ddagger$] &  $5.0\pm1.1$
	&  $0.063$\footnotemark[$\ddagger$] \\
NGC~5195  & (\timeform{0"},   \timeform{0"})
	& $<0.13\pm0.06$\footnotemark[$\S$]       &  $3.2\pm0.6$
	&  $0.018\pm0.007$\footnotemark[$\S$] \\
\hline
\multicolumn{5}{@{}l@{}}{\hbox to 0pt{\parbox{110mm}{\footnotesize
	Notes.
	\par\noindent
	\footnotemark[$*$] Offset from the position of the nucleus
		indicated in Tables~\ref{table-sample-sb} and
		\ref{table-sample-npb}.
	\par\noindent
	\footnotemark[$\dagger$] Ratio of integrated intensity indicated
		in Table~\ref{table-res-prop}.
		Integrated intensity that is not in
		Table~\ref{table-res-prop} is taken from previously published
		papers, as indicated below.
		Quoted uncertainties are 1$\sigma$, and upper limits are
		2$\sigma$, except for the HCN/$^{13}$CO value of NGC~5195
		(see note $\S$ below).
	\par\noindent
	\footnotemark[$\ddagger$] HCN(1 -- 0) integrated intensity and
		HCN/$^{12}$CO ratio taken from \citet{koh98}.
	\par\noindent
	\footnotemark[$\S$] HCN(1 -- 0) integrated intensity and
		HCN/$^{12}$CO ratio taken from \citet{koh02}.
		Due to the off-position problem, we take HCN/$^{13}$CO as
		an upper limit (see Sect.\ref{sect-res} for detail).
	\par\noindent
	\footnotemark[$\|$] $^{12}$CO integrated intensity taken from
		\citet{sor00}.
	\par\noindent
	\footnotemark[$\#$] $^{12}$CO integrated intensity taken from
		\citet{sor02}.
	\par\noindent
	\footnotemark[$**$] $^{12}$CO integrated intensity taken from
		\citet{kun07}.
	\par\noindent
	\footnotemark[$\dagger\dagger$] $^{12}$CO integrated intensity
		taken from \citet{kun00}.
	}\hss}}
\end{tabular}
\end{center}
\end{table*}

\begin{figure}
\begin{center}
\FigureFile(80mm,160mm){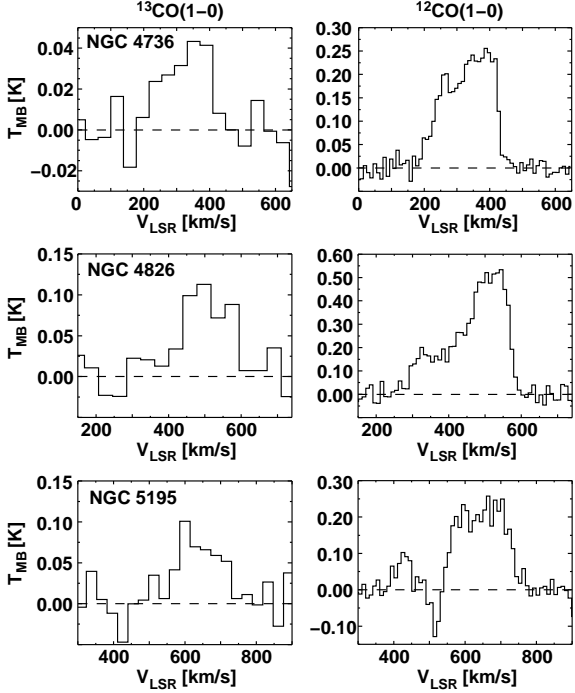}
\end{center}
\caption{$^{13}$CO(1 -- 0) and $^{12}$CO(1 -- 0) spectra toward the
	center of non-starburst galaxy samples.
	Two spectra are shown for each galaxy; the spectrum on the
	left-hand side denotes the $^{13}$CO(1 -- 0) line, and that on
	the right-hand side denotes the $^{12}$CO(1 -- 0) line.
	The horizontal and vertical axes are $V_{\rm LSR}$ and
	$T_{\rm MB}$, respectively.
\label{fig-res-nsb}}
\end{figure}

Fig.~\ref{fig-res-nsb} shows the $^{13}$CO(1 -- 0) and
$^{12}$CO(1 -- 0) line spectra of the non-starburst galaxy samples.
The integrated intensity of $^{12}$CO(1 -- 0) line, I($^{12}$CO), for
NGC~4736 is consistent with that of \citet{koh98}.
For I($^{12}$CO) of NGC~4826, it is 67\% of the value obtained by
\citet{koh98}.
This disagreement may be explained by pointing error and/or
calibration error.
The pointing error of our observations was $\sim4''$, larger than the
$\sim2''$ of \citet{koh98}.
On the other hand, \citet{koh98} did not observe any intensity
calibrator as we did with IRC$+$10216.
For I($^{12}$CO) of NGC~5195, we only detect 46\% of the value
obtained by \citet{koh02}.
This is because we had mistakenly placed the off-position of the
position switching observations at NGC~5194, which is the counterpart
of the M51 interaction system for NGC~5195 and rich in molecular gas.
This can be seen as the absorption-like feature around the velocity
range of $500-550$~km~s$^{-1}$ in Fig.~\ref{fig-res-nsb}, which did
not show up in the spectrum of \citet{koh02}.
Therefore we use the derived HCN/$^{13}$CO value for NGC~5195 as an
upper limit.

The HCN/$^{13}$CO integrated intensity ratios between the starburst
and non-starburst galaxy samples are significantly different:
We obtained an average HCN/$^{13}$CO toward the nuclei of the
starburst samples of $1.15\pm0.32$, and a value of $1.10\pm0.30$ for
starburst samples including the off-center starburst regions of M82,
namely the NE and SW lobes.
The average HCN/$^{13}$CO ratio toward the nuclei of the
non-starburst galaxy samples and that including the non-starburst
regions of M82, namely (\timeform{-30"}, \timeform{-6"}) and
(\timeform{-45"}, \timeform{-9"}), are $<0.31\pm0.14$ and
$<0.32\pm0.12$, respectively.
These values are consistent with previously published results:
Paper I suggests that starburst regions tend to have higher
HCN/$^{13}$CO of $>1$, but less active star forming regions tend to
have lower ratios ($\lesssim1$).
Note that the results of Paper I is based on interferometric
observations; therefore it is sensitive to compact components, such
as high density gas (i.e., HCN).
It follows that ratios in Paper I are bound to take on larger values
than those in this paper, which is based on single-dish observations.

The $^{12}$CO/$^{13}$CO and HCN/$^{12}$CO integrated intensity ratios
between starburst and non-starburst galaxy samples also show
differences with the average values for the starburst samples larger
than those for the non-starburst samples:
The average $^{12}$CO/$^{13}$CO integrated intensity ratios
toward the nuclei of the starburst and non-starburst samples are
$10.2\pm3.2$ and $5.1\pm1.6$, respectively, and the average
HCN/$^{12}$CO integrated intensity ratios for these samples are
$0.11\pm0.04$ and $<0.04\pm0.02$, respectively.

All the HCN/$^{13}$CO, $^{12}$CO/$^{13}$CO, and HCN/$^{12}$CO ratios
for the starburst galaxy samples are $2-3$ times larger than those for
the non-starburst galaxy samples.
These results suggest that the conditions of molecular gas in the
starburst regions differ significantly from those in the
non-starburst regions.

\section{Discussions}
\label{sect-dis}

\subsection{Physical Conditions of Molecular Gas in Sample Galaxies}
\label{sect-dis-phys}

Using the line ratios derived above, we can estimate the physical
conditions of molecular gas in our starburst and non-starburst sample
galaxies.
To that end, we make use of the large-velocity-gradient (LVG)
approximation \citep{gol74,sco74} assuming a one-zone model, and the
photon dominated region (PDR) model \citep{mei05,mei07}.
Note that the estimated physical conditions are averaged values
within one beam size of our observations, which corresponds to a few
hundred to a few kpc scale in linear size.
Although deriving an averaged value within one beam is extremely
simplified assumption compared to the real situations of molecular
clouds and radiation fields in galaxies, it is successfully derived
in various galaxies in the past studies.
The overlapping effect of molecular clouds is expected to be small,
based on the small shadowing effect in molecular clouds of our Galaxy
that is edge-on \citep{bur78}, so we assume that we see most of the
emission from each molecular clouds in our beam, and therefore it is
safe to use the LVG and PDR models in our datasets.

\citet{spe92} suggested that the increase of the HCN intensity toward
the centers of galaxies can be due to the increase of gas pressure,
not due to the increase of gas density induced by gravitational
instability.
Hence, they predicted that larger galaxies should have stronger HCN
intensities than those in smaller galaxies.
This prediction can be rejected from the comparison of the
HCN/$^{12}$CO and HCN/$^{13}$CO ratios between our Galaxy and M82:
Our Galaxy, which is a larger galaxy, has the HCN/$^{12}$CO and
HCN/$^{13}$CO ratios of $\sim0.08$ \citep{jac96,hel97} and $\sim0.8$
\citep{mat98}, respectively, at the Galactic center.
The center of a much smaller galaxy, M82, has much larger
HCN/$^{12}$CO and HCN/$^{13}$CO ratios of $\sim0.1$ and $\sim1.6$,
respectively.
Similar conclusion is also obtained by \citet{gao04b}.
We therefore assume that the differences of ratios between galaxies
and within galaxies are due to the differences in physical conditions
of molecular gas, not due to the differences in gas pressure.

\subsubsection{LVG Model}
\label{sect-dis-lvg}

\begin{figure*}
\begin{center}
\FigureFile(170mm,80mm){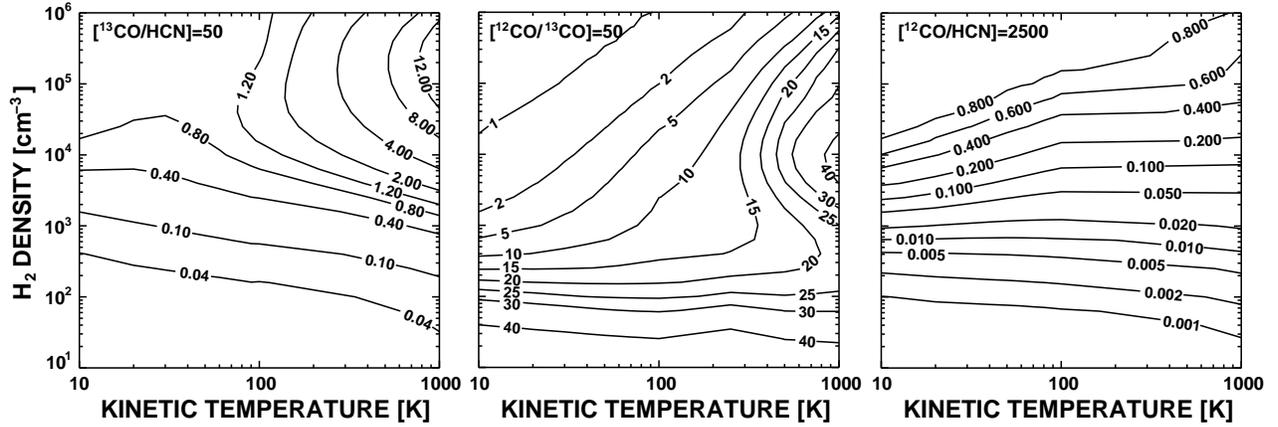}
\end{center}
\caption{H$_{2}$ number density (vertical axis) and kinetic
	temperature (horizontal axis) dependence of the
	HCN(1 -- 0)/$^{13}$CO(1 -- 0) (left),
	$^{12}$CO(1 -- 0)/$^{13}$CO(1 -- 0) (middle), and
	HCN(1 -- 0)/$^{12}$CO(1 -- 0) (right) intensity ratios derived
	using the LVG calculations.
	Refer to the main text for the details of the calculations.
\label{fig-dis-lvg}}
\end{figure*}

The line intensity ratios for CO and HCN molecules under the LVG
model were tabulated as a function of the kinetic temperatures,
$T_{\rm k}$, from 10~K to 1000~K, and the H$_{2}$ number density,
$n({\rm H_{2}})$, from $10^{1}$~cm$^{-3}$ to $10^{6}$~cm$^{-3}$ for
HCN(1 -- 0)/$^{13}$CO(1 -- 0), $^{12}$CO(1 -- 0)/$^{13}$CO(1 -- 0),
and HCN(1 -- 0)/$^{12}$CO(1 -- 0).
The collision rates for CO molecules $\leq250$~K and $\geq500$~K were
taken from \citet{flo85} and \citet{mck82}, respectively.
The corresponding values for HCN molecules $<100$~K and $\geq100$~K
are available from \citet{gre74} and on-line\footnotemark[$*$].
\footnotetext[$*$]{The data are available at \\
 http://data.giss.nasa.gov/mcrates/data/hcn\_he\_rates.txt.}
We fixed the abundances to the `standard' relative abundance where
$Z(^{13}{\rm CO})$ = [$^{13}$CO]/[H$_{2}$] $=1\times10^{-6}$
\citep{sol79}, $Z({\rm HCN})$ = [HCN]/[H$_{2}$] $=2\times 10^{-8}$
\citep{irv87}, and [$^{12}$CO]/[$^{13}$CO] = 50.
We also set the velocity gradient, $dv/dr$, to a fixed general value
of 1.0~km~s$^{-1}$~pc$^{-1}$.
The tabulated results are shown in Fig.~\ref{fig-dis-lvg}.

The LVG calculations show that high HCN/$^{13}$CO ratio ($\gtrsim1$)
can be obtained under high density
($n({\rm H_{2}})\sim10^{5\pm1}$~cm$^{-3}$) and high temperature
($T_{\rm k}\gtrsim50$~K) conditions.
The calculations for $^{12}$CO/$^{13}$CO show that high ratio ($>10$)
can be obtained under high density
($n({\rm H_{2}})\sim10^{4\pm1}$~cm$^{-3}$) and high temperature
($T_{\rm k}>100$~K) conditions, or low density
($n({\rm H_{2}})\lesssim10^{2}$~cm$^{-3}$) conditions.
The HCN/$^{12}$CO ratio, on the other hand, only traces the density,
and a high ratio ($\gtrsim0.1$) indicates high density
($n({\rm H_{2}})>2\times10^{3}$~cm$^{-3}$) condition.
These results can be explained as follows:
At high temperature, the energy distribution of the $^{13}$CO
rotational transition populating higher levels due to the low
abundance and small dipole moment of the $^{13}$CO molecules, and
therefore the $^{13}$CO(1 -- 0) line will be optically thin with a
corresponding decrease in the line intensity.
On the other hand, $^{12}$CO molecules are abundant and HCN molecules
possess large dipole moments, therefore the $^{12}$CO(1 -- 0) and
HCN(1 -- 0) lines remain optically thick even under high temperature
conditions.
In terms of density, the HCN molecule has a high critical density,
meaning that high density is a prerequisite condition for line
emission.
The $^{12}$CO and $^{13}$CO molecules have the same critical density
but $^{12}$CO emits under lower density conditions due to its
abundance.

$^{13}$CO deficiency is often cited in the case of merging galaxies
(i.e., extreme starburst galaxies; e.g., \cite{aal91,cas92}).
Our sample galaxies do not exhibit such extreme starburst behaviors,
but the possibility of low $Z(^{13}{\rm CO})$ cannot be ruled out.
In addition, starburst galaxies may have larger velocity gradients
due to the large turbulences caused by active star formation.
In this study, we checked the dependence of $Z(^{13}{\rm CO})$ and
the velocity gradient with our LVG calculation results:
If $Z(^{13}{\rm CO})$ is a factor of two lower than the standard
abundance, or if the velocity gradient is a factor of two higher than
the standard value we assumed above, the LVG calculations for
$^{12}$CO/$^{13}$CO and HCN/$^{13}$CO give rise to a temperature a
factor of two lower.

\begin{table}
\begin{center}
\caption{Physical conditions of molecular gas in the sample galaxies
	based LVG calculations.}
\label{table-dis-phys}
\begin{tabular}{lccc}
\hline
Galaxy & Position\footnotemark[$*$]
	& $n({\rm H}_{2})$\footnotemark[$\dagger$]
	& $T_{\rm k}$\footnotemark[$\ddagger$] \\
\hline
NGC~253   & (\timeform{0"},   \timeform{0"})  & $\sim(1-2)\times10^{4}$
	&   $100-200$  \\
NGC~2146  & (\timeform{0"},   \timeform{0"})  & $\sim(3-5)\times10^{3}$
	&   $100-200$  \\
NGC~2903  & (\timeform{0"},   \timeform{0"})  & $\sim(6-9)\times10^{3}$
	&   $100-200$  \\
M82       & (\timeform{0"},   \timeform{0"})  & $\sim(6-9)\times10^{3}$
	&   $200-300$  \\
(NE lobe) & (\timeform{+10"}, \timeform{+6"}) & $\sim4\times10^{4\pm1}$
    &  $\gtrsim50$ \\
(SW lobe) & (\timeform{-15"}, \timeform{-3"}) & $\sim(3-5)\times10^{3}$
    &   $\sim200$  \\
          & (\timeform{-30"}, \timeform{-6"}) & $\sim(2-3)\times10^{3}$
    &   $\sim100$  \\
          & (\timeform{-45"}, \timeform{-9"}) & $\lesssim5\times10^{3}$
    &      ---     \\
NGC~3504  & (\timeform{0"},   \timeform{0"})  & $\sim1\times10^{4}$
	&   $\sim200$  \\
NGC~6946  & (\timeform{0"},   \timeform{0"})  & $\sim(6-9)\times10^{3}$
	&   $100-200$  \\
NGC~6951  & (\timeform{0"},   \timeform{0"})  & $\sim(6-9)\times10^{3}$
	&    $60-100$  \\
\hline
NGC~4736  & (\timeform{0"},   \timeform{0"})  & $\lesssim2\times10^{3}$
	& $\lesssim60$ \\
NGC~4826  & (\timeform{0"},   \timeform{0"})  & $\sim(3-4)\times10^{3}$
	&  $\sim50-60$ \\
NGC~5195  & (\timeform{0"},   \timeform{0"})  &     $\sim1\times10^{3}$
	&  $\sim10-20$ \\
\hline
\multicolumn{4}{@{}l@{}}{\hbox to 0pt{\parbox{90mm}{\footnotesize
	Notes.
	\par\noindent
	\footnotemark[$*$] Offset from the position of the nucleus
		indicated in Tables~\ref{table-sample-sb} and
		\ref{table-sample-npb}.
	\par\noindent
	\footnotemark[$\dagger$] Molecular hydrogen number density in
		cm$^{-3}$.
	\par\noindent
	\footnotemark[$\ddagger$] Kinetic temperature in K.
	}\hss}}
\end{tabular}
\end{center}
\end{table}

Using the LVG calculations shown in the figure, we arrived at the
physical conditions of molecular gas in the sample galaxies and
summarized them in Table~\ref{table-dis-phys}.
Many of the starburst galaxy samples display high density
($>5\times10^{3}$~cm$^{-3}$) and high temperature ($>100$~K).
But some of the galaxies (NGC 2146 and NGC 6951) have been estimated
as rather lower density or temperature due to lower HCN/$^{13}$CO
values, even these galaxies have high infrared luminosities (see
Table~\ref{table-sample-sb}).
This is probably because of the larger distances of these galaxies
than other nearby galaxies: Due to larger distances, the beam
observes larger radius, where usually dominated by low density and
low temperature materials, than other nearby galaxies, and therefore
the beam averaged physical conditions exhibit lower values than other
nearby galaxies.
On the other hand, the non-starburst galaxy samples displays lower
densities and temperatures of
$n({\rm H_{2}})\lesssim4\times10^{3}$~cm$^{-3}$ and
$T_{\rm k}\lesssim60$~K than the starburst galaxy samples,
even non-starburst samples located nearby.

We then compared our estimations with previously published values.
In the case of M82, a region of higher density
($\sim4\times10^{4\pm1}$~cm$^{-3}$) can be observed at the NE lobe,
which gradually tapers off toward the southwestern side of the
galaxy.
This trend, as well as the corresponding derived values, are very
similar to those derived by \citet{pet00}.
The temperature, on the other hand, has high values ($200-300$~K) at
the nucleus and the SW lobe, and $\sim100$~K at the position
(\timeform{-30"}, \timeform{-6"}).
The decrease of the temperature as a function of radius in M82 (and
in other galaxies) is also suggested from the $^{12}$CO/$^{13}$CO
observations \citep{pag01}.
The temperature of the NW lobe is not well constrained,
$\gtrsim50$~K, and there exists no constraint for the position
(\timeform{-45"}, \timeform{-9"}) due to the absence of
$^{12}$CO(1 -- 0) data.
These estimated temperatures and densities tend to be higher than
those derived using various CO lines (including isotopes) observed
with an interferometer \citep{wei01}.

Estimations for NGC~253 yield
$n({\rm H_{2}})\sim(1-2)\times10^{4}$~cm$^{-3}$ and
$T_{\rm k}\sim100-200$~K, which are very similar to previously
published results, utilizing several molecular species
\citep{wal91,pag95,bra03,bay04}, but rather higher than those derived
using only $^{12}$CO lines \citep{gus06}.
For NGC~6946, our estimations suggest that
$n({\rm H_{2}})\sim(6-9)\times10^{3}$~cm$^{-3}$ and
$T_{\rm k}\sim100-200$~K, which are similar to the values obtained by
\citet{mei04} using various CO lines including isotopes and also
HCN(1-0) line, but rather higher than those derived by \citet{wal02},
which only made use of the CO lines (including isotopes).

\begin{table*}[t]
\begin{center}
\footnotesize
\caption{IRAS data and infrared flux ratios of sample galaxies.}
\label{table-dis-iras}
\begin{tabular}{ccccccc}
\hline
Galaxy
         & $\log\frac{f(12~\micron)}{f(25~\micron)}$
         & $\log\frac{f(12~\micron)}{f(60~\micron)}$
         & $\log\frac{f(12~\micron)}{f(100~\micron)}$
         & $\log\frac{f(25~\micron)}{f(60~\micron)}$
         & $\log\frac{f(25~\micron)}{f(100~\micron)}$
         & $\log\frac{f(60~\micron)}{f(100~\micron)}$ \\
\hline
NGC~253
         & $-0.5762\pm0.0004$     & $-1.3726\pm0.0001$
         & $-1.4968\pm0.0004$     & $-0.7964\pm0.0001$
         & $-0.9206\pm0.0003$     & $-0.1242\pm0.0002$     \\
NGC~2146
         & $-0.4400\pm0.0016$     & $-1.3320\pm0.0005$
         & $-1.4535\pm0.0015$     & $-0.8920\pm0.0007$
         & $-1.0135\pm0.0009$     & $-0.1215\pm0.0006$     \\
NGC~2903
         & $-0.2131\pm0.0045$     & $-1.0586\pm0.0018$
         & $-1.3919\pm0.0029$     & $-0.8455\pm0.0035$
         & $-1.1789\pm0.0035$     & $-0.3333\pm0.0008$     \\
M82
         & $-0.6220\pm0.0002$     & $-1.2704\pm0.00005$
         & $-1.2379\pm0.0003$     & $-0.6484\pm0.00007$
         & $-0.6159\pm0.0002$     & $+0.0325\pm0.0002$    \\
NGC~3504
         & $-0.5600\pm0.0125$     & $-1.2857\pm0.0034$
         & $-1.4868\pm0.0116$     & $-0.7257\pm0.0049$
         & $-0.9268\pm0.0049$     & $-0.2011\pm0.0015$     \\
NGC~6946
         & $-0.2328\pm0.0013$     & $-1.0301\pm0.0007$
         & $-1.3803\pm0.0013$     & $-0.7972\pm0.0007$
         & $-1.1475\pm0.0009$     & $-0.3502\pm0.0007$     \\
NGC~6951
         & $-0.2073\pm0.0053$     & $-1.0835\pm0.0029$
         & $-1.4938\pm0.0049$     & $-0.8761\pm0.0029$
         & $-1.2864\pm0.0034$     & $-0.4103\pm0.0020$     \\
\hline
NGC~4736
         & $-0.0810\pm0.0055$     & $-1.1495\pm0.0040$
         & $-1.3767\pm0.0048$     & $-1.0685\pm0.0029$
         & $-1.2956\pm0.0029$     & $-0.2271\pm0.0009$     \\
NGC~4826
         & $-0.0835\pm0.0105$     & $-1.1918\pm0.0048$
         & $-1.5390\pm0.0057$     & $-1.1083\pm0.0089$
         & $-1.4556\pm0.0089$     & $-0.3473\pm0.0011$     \\
NGC~5195\footnotemark[$\dagger$]
         & $-0.4645:$             & $-1.6383:$
         & $-1.9512:$             & $-1.1738:$
         & $-1.4874:$             & $-0.3135:\pm0.0228$    \\
\hline
\multicolumn{7}{@{}l@{}}{\hbox to 0pt{\parbox{180mm}{\footnotesize
	Notes.
	\par\noindent
	Flux data are taken from \citet{san03}.
	\par\noindent
	\footnotemark[$\dagger$] A value with a colon (``:'') indicates
		that the value has a large uncertainty.
	}\hss}}
\end{tabular}
\end{center}
\end{table*}

The above comparisons indicate that our approach tends to yield
higher densities and temperatures than other studies using only the
$^{12}$CO lines or even the CO isotopes.
This is because our methodology makes extensive use of both the high
density tracer HCN(1-0) line and the optically thin and therefore
temperature dependent tracer $^{13}$CO(1-0) line.
In particular, HCN(1-0)/$^{13}$CO(1-0) is sensitive to both density
and temperature, once it attains a value of around unity or higher
(see Paper I for more detailed discussions).
We therefore concluded that the combinations of these three line
ratios, or even HCN(1-0)/$^{13}$CO(1-0) by itself, can effectively
distinguish galaxies with warm and dense molecular gas from galaxies
without.

\subsubsection{PDR Model}
\label{sect-dis-pdr}

We also calculated HCN(1-0)/$^{13}$CO(1-0) with the PDR model.
It turns out that the PDR model can also yield a HCN/$^{13}$CO of
around unity; it requires an interstellar radiation field (ISRF)
strength of around several $\times10^{3}$~G$_{0}$ and a density of
around several $\times10^{5}$~cm$^{-3}$.
In other words, dense molecular gas irradiated by radiation from
massive stars can reproduce a high HCN/$^{13}$CO of around unity.
The ratio decreases with decreasing ISRF strength or density, but
the decrease of the ratio is sensitive to the density decrease than
to the ISRF strength decrease.

Using this PDR model, the ISRF strength and the density for the
starburst sample are estimated as around $10^{2-3}$~G$_{0}$ and on
the order of $10^{5}$~cm$^{-3}$, respectively.
The non-starburst sample, on the other hand, has lower ISRF of the
order of $10^{2}$~G$_{0}$ and lower density of the order of
$10^{4}$~cm$^{-3}$ or less.

The strength of the ISRF can be derived also from the IR
characteristics.
The range of the ISRF strength in nearby galaxies is
$\sim10^{2-4}$~G$_{0}$ \citep{neg01,mal01}, and exhibits a tight
correlation with the IR flux ratios between $60~\mu$m and $100~\mu$m,
$f(60~\micron)/f(100~\micron)$; higher the ratio, higher the ISRF
strength \citep{neg01}.
Our sample galaxies (both starburst and non-starburst samples) have
a wide range of $f(60~\micron)/f(100~\micron)$, and our starburst
sample has a tendency of higher ratio than that of the non-starburst
sample, although the range is overlapping with the non-starburst
galaxy sample (see Fig.~\ref{fig-sample-ir} and
Table~\ref{table-dis-iras}).
If we adopt the correlation shown by \citet{neg01}, our starburst
sample has higher ISRF strength of the order of $10^{2-3}$~G$_{0}$
than the non-starburst sample of the order of $10^{2}$~G$_{0}$.
Those values are consistent with the values derived from
HCN/$^{13}$CO using the PDR model, suggesting that there is a
correlation between HCN/$^{13}$CO and $f(60~\micron)/f(100~\micron)$.
In the next subsection, we study this relation, and the relations
between HCN/$^{13}$CO and other IR flux ratios.

\subsection{Relation between Molecular Gas and Dust}
\label{sect-dis-rel}

Dust in galaxies absorbs the radiation from stars, in particular
massive stars, and re-radiate at IR wavelengths.
The IR emission from galaxies, especially IR flux ratios, is
therefore a good tracer for star formation activities
(e.g., \cite{hel86}; see also reviews of \cite{soi87} and
 \cite{tel93}).
The spectral energy distribution study using the IRAS and the
Infrared Space Observatory (ISO) data between 3 and 1100~$\mu$m
further confirmed the utility of the IR flux ratios for tracing
star formation activities, and also suggested that the mid-IR
continuum flux at a range of 20 -- 42~$\mu$m can be the optimum dust
emission tracer for current star formation activities \citep{dal01}.
Here we compare HCN/$^{13}$CO and IR continuum flux ratios to
ascertain the relation between molecular gas and dust properties.

In this study, we use the IRAS flux \citep{san03} for the infrared
flux due to the following reasons:
(1) The IRAS data are still the only all-sky survey results so far,
	and all the wavelength data exist for all our sample galaxies.
(2) Although the beam size of the IRAS data is large (arcminutes
	scale), most far-IR radiation from nearby galaxies is
	concentrated in the central $\sim30''$ \citep{smi96}.
	The regions that traced by our observations and that by the IRAS
	are therefore more or less similar.
	Indeed, previously published single dish molecular gas studies
	derived important results using the IRAS data.
The IRAS flux ratios for our sample galaxies are tabulated in
Table~\ref{table-dis-iras}.

\begin{figure*}
\begin{center}
\FigureFile(170mm,170mm){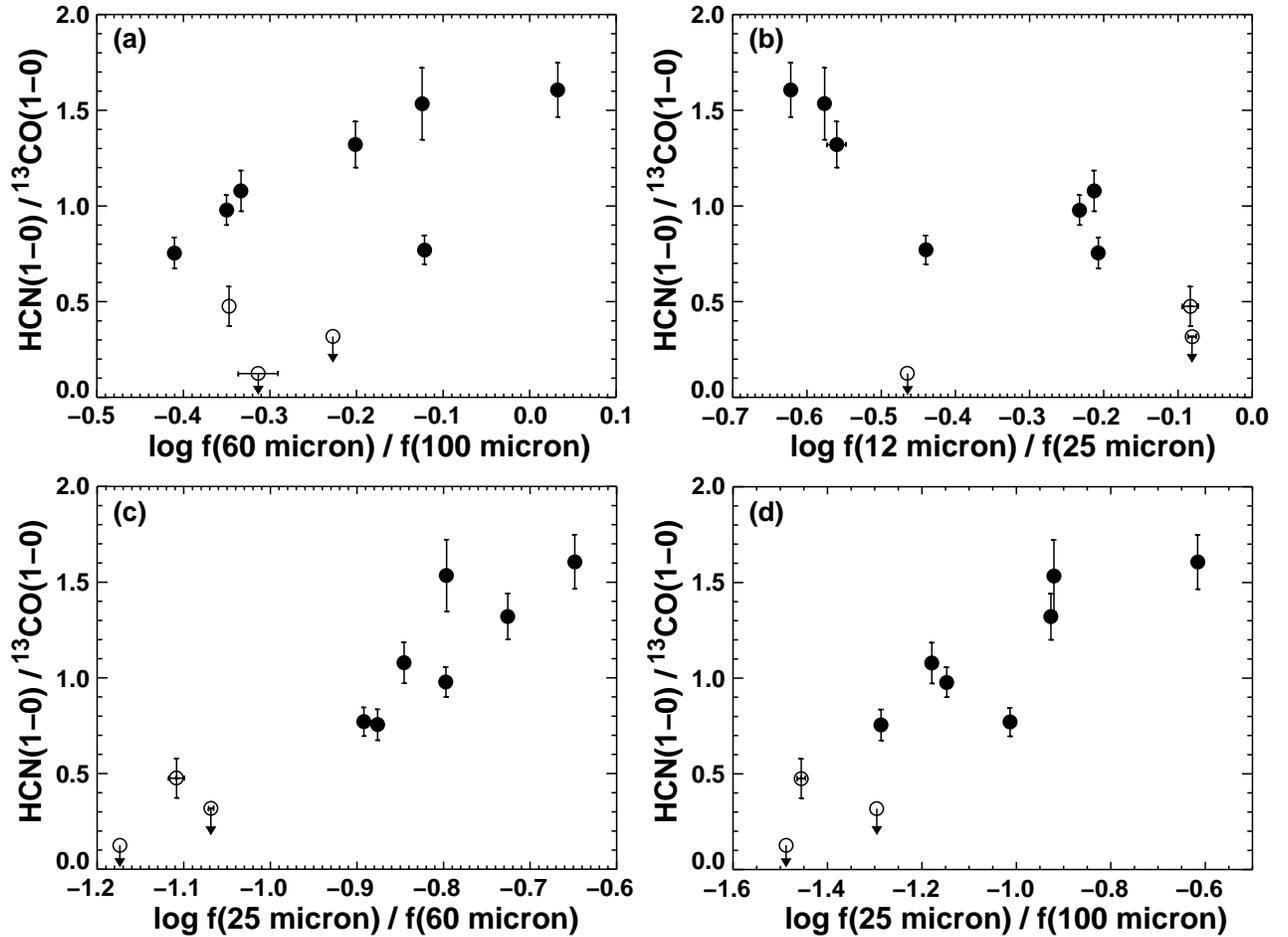}
\end{center}
\caption{Correlation diagrams between the
	HCN(1 -- 0)/$^{13}$CO(1 -- 0) integrated intensity ratios and
	(a) $f(60~\micron)/f(100~\micron)$,
	(b) $f(12~\micron)/f(25~\micron)$,
	(c) $f(25~\micron)/f(60~\micron)$, and 
	(d) $f(25~\micron)/f(100~\micron)$ IR flux ratios.
	Filled and open circles are the same as in Fig.~\ref{fig-sample-ir}.
	Downward arrows denote upper limits.
	All the diagrams, especially (b), (c), and (d), exhibit tight
	correlations.
\label{fig-dis-corr}}
\end{figure*}

Since $f(12~\micron)/f(25~\micron)$ and
$f(60~\micron)/f(100~\micron)$ are good tracers of star formation
(see Sect.~\ref{sect-sample} and Fig.~\ref{fig-sample-ir}), and the
relation between HCN/$^{13}$CO and $f(60~\micron)/f(100~\micron)$ is
suggested in the previous subsection, we first compare HCN/$^{13}$CO
with these IR ratios (Fig.~\ref{fig-dis-corr}a, b).
HCN/$^{13}$CO indeed exhibits a correlation with
$f(60~\micron)/f(100~\micron)$.
There exists also a correlation between HCN/$^{13}$CO and
$f(12~\micron)/f(25~\micron)$ ratio, which is natural since
$f(12~\micron)/f(25~\micron)$ correlates with
$f(60~\micron)/f(100~\micron)$ \citep{hel86,dal01}.
In addition, the correlation is much tighter than the case with
$f(60~\micron)/f(100~\micron)$, except for NGC~5195 that has a large
uncertainty in the IR flux (see Table~\ref{table-dis-iras}).

In Fig.~\ref{fig-dis-corr}(c) and (d), we also included correlation
diagrams for $f(25~\micron)/f(60~\micron)$ and
$f(25~\micron)/f(100~\micron)$, and both show tight correlations as
seen in the HCN/$^{13}$CO -- $f(12~\micron)/f(25~\micron)$ diagram.
On the other hand, there is no significant correlation with other
IR flux ratios, namely $f(12~\micron)/f(60~\micron)$ and
$f(12~\micron)/f(100~\micron)$ (diagrams not shown).

The tight correlations between HCN/$^{13}$CO and the $25~\mu$m
related IR flux ratios indicate that the increase of the $25~\mu$m
continuum flux relative to the other IR continuum bands is tightly
correlated with the increase of HCN/$^{13}$CO.
Under the LVG model, HCN/$^{13}$CO is sensitive to molecular gas that
has a temperature of 100~K or more.
Under the PDR model, HCN/$^{13}$CO is sensitive to the molecular gas
density and the strength of the ISRF.
The increase of the $25~\mu$m continuum flux is mainly due to the
increase of flux from very small grains \citep{dal01}.
It is believed that the heating of these grains is intermediate
between thermal equilibrium and single-photon heating, and the
increase in flux is closely related to the increase in the heating
intensity of the ISRF \citep{dal01}.
This is consistent with the results from the PDR model.
Furthermore, the black body temperature around $25~\mu$m is around
100~K or more, therefore dust needs to be heated up to approximately
this temperature before achieving thermal equilibrium, in complete
agreement with the results from the LVG model.
The tight correlations therefore suggest that the molecular gas and
dust are localized in the same regions, with the strong radiation
from starburst activities heating the dust and in turn the molecular
gas.
With the dust heated up to around 100~K or more, the molecular gas
will follow suit.
In either LVG or PDR model, this tight correlation suggests that
HCN/$^{13}$CO will be an excellent tracer for starburst conditions.
It also applies that either or both the LVG and PDR mechanisms play
important roles in star forming regions.

It is now well established that the amount of dense molecular gas is
closely correlated to star formation activities
\citep{sol92,gao04a,gao04b}.
Our results suggests that in addition to the amount of dense
molecular gas, the amount of dense {\it and warm} molecular gas is
also pertinent to star formation activities.
The possible causes of this correlation are the heating of molecular
gas due to strong radiation and stellar wind from newly formed
massive stars, and/or numerous supernova explosions.

\section{Conclusion}
\label{sect-concl}

We conducted a survey of $^{12}$CO(1 -- 0), $^{13}$CO(1 -- 0), and
HCN(1 -- 0) lines toward seven starburst and three non-starburst
galaxies nearby using the Nobeyema 45~m telescope (beam size
$\sim14\arcsec-18\arcsec$).
The $^{13}$CO(1 -- 0) and HCN(1 -- 0) lines were obtained for all the
sample galaxies, but the $^{12}$CO(1 -- 0) data for starburst galaxy
samples and the HCN data for non-starburst galaxy samples were
obtained from previously published data.
\begin{itemize}
\item The $^{13}$CO(1 -- 0) and HCN(1 -- 0) intensities in the
	central regions of the starburst galaxy samples are more or less
	similar, with the HCN/$^{13}$CO integrated intensity ratios
	ranging between 0.75 -- 1.61 (average $=1.15\pm0.32$).
	The $^{13}$CO(1 -- 0) intensities in the non-starburst galaxy
	samples are always stronger than the HCN(1 -- 0) intensities, and
	HCN/$^{13}$CO yield obviously smaller values of $<0.31\pm0.14$
	than for the starburst galaxy samples.
	We observed multiple positions for the starburst galaxy M82, and
	the ratios far from the center (i.e., disk region) are similar to
	those in the non-starburst galaxy samples.
\item $^{12}$CO/$^{13}$CO and HCN/$^{12}$CO also show higher values
	for about a factor of two in starburst galaxy samples than those
	in the non-starburst galaxy sample.
\item The physical attributes of molecular gas derived from the line
	ratios under the LVG model indicate dense
	($>5\times10^{3}$~cm$^{-3}$) and warm ($\gtrsim100$~K) conditions
	in most of the central regions of the starburst galaxy samples,
	but diffuse ($\lesssim4\times10^{3}$~cm$^{-3}$) and cold
	($\lesssim60$~K) conditions in the non-starburst galaxy samples.
	Under the PDR model, this suggests that the dense molecular gas
	($\sim10^{5}$~cm$^{-3}$) is being irradiated by ISRF of
	$10^{2-3}$~G$_{0}$ in the starburst galaxy samples, but lower
	density molecular gas of $\lesssim10^{4}$~cm$^{-3}$ is being
	irradiated by lower intensity ISRF of $10^{2}$~G$_{0}$ in the
	non-starburst galaxy samples.
\item HCN/$^{13}$CO exhibits tight correlations with the IRAS flux
	ratios between $25~\micron$ and other wave bands, namely, the
	increase of HCN/$^{13}$CO is related to the increase of the IRAS
	$25~\micron$ flux.
	Since the IRAS $25~\micron$ flux is a good tracer for star
	formation activities, these correlations indicate that
	HCN/$^{13}$CO is also a good tracer for star formation
	activities.
	High HCN/$^{13}$CO larger than unity points to dense and warm
	molecular gas with the temperature of 100~K or more under the LVG
	model, or irradiated by strong ISRF under the PDR model.
	The high $25~\micron$ flux also indicates high ISRF and high dust
	temperature of around 100~K or more.
	These results suggest that molecular gas and dust are localized
	in the same regions in a well mixed configuration, and that the
	strong radiation heats up the dust and therefore the molecular
	gas nearby indirectly.
	This strong radiation also affects the molecular gas itself.
\end{itemize}

\bigskip

We are grateful to the NRO staff for the operation and improvement of
the Nobeyama 45~m telescope.
This work is supported by the National Science Council (NSC) of
Taiwan, NSC 96-2112-M-001-009 and NSC 97-2112-M-001-021-MY3.

\end{document}